\documentclass[aps,prl,floatfix,twocolumn,showpacs,superscriptaddress,groupedaddress]{revtex4-1}  
\usepackage{graphicx}  
\usepackage{amsmath,amssymb}
\usepackage{flushend}
\usepackage{array}
\usepackage{float}
\usepackage{natbib}
\usepackage{booktabs}
\usepackage{color}

\begin{document}

\title{Electrically switchable giant Berry curvature dipole in silicene, germanene and stanene}

\author{Arka Bandyopadhyay} \affiliation{Solid State and Structural Chemistry Unit,
Indian Institute of Science, Bangalore 560012, India.}
\author{Nesta Benno Joseph}\affiliation{Solid State and Structural Chemistry Unit,
Indian Institute of Science, Bangalore 560012, India.}
\author{Awadhesh Narayan} \email[]{awadhesh@iisc.ac.in} \affiliation{Solid State and Structural Chemistry Unit,
Indian Institute of Science, Bangalore 560012, India.}

\vskip 0.25cm

\date{\today}

\begin{abstract}
The anomalous Hall effect in time-reversal symmetry broken systems is underpinned by the concept of Berry curvature in band theory. However, recent experiments reveal that the nonlinear Hall effect can be observed in non-magnetic systems without applying an external magnetic field. The emergence of nonlinear Hall effect under time-reversal symmetric conditions can be explained in terms of non-vanishing Berry curvature dipole arising from inversion symmetry breaking. In this work, we availed realistic tight-binding models, first-principles calculations, and symmetry analyses to explore the combined effect of transverse electric field and strain, which leads to a giant Berry curvature dipole in the elemental buckled honeycomb lattices -- silicene, germanene, and stanene. The external electric field breaks the inversion symmetry of these systems, while strain helps to attain an asymmetrical distribution of Berry curvature of a single valley. Furthermore, the topology of the electronic wavefunction switches from the band inverted quantum spin Hall state to normal insulating one at the gapless point. This band gap closing at the critical electric field strength is accompanied by an enhanced Berry curvature and concomitantly a giant Berry curvature dipole at the Fermi level. Our results predict the occurrence of an electrically switchable nonlinear electrical and thermal Hall effect in a new class of elemental systems that can be experimentally verified.
\end{abstract}

\maketitle


\section{Introduction}

The appearance of Hall current is invariably contingent on the breaking of time-reversal symmetry (TRS) in the linear response regime \cite{klitzing1980new}. TRS in a material can be broken by an external magnetic field, or suitable magnetic dopants. However, in recent experiments, \emph{nonlinear} Hall effects (NHE) \cite{du2021nonlinear,ortix2021nonlinear} have been detected in non-magnetic transition-metal dichalcogenides (TMDs) under time-reversal-symmetric conditions \cite{ma2019observation,kang2019nonlinear,huang2020giant}. In their seminal work, Sodemann and Fu \cite{sodemann2015quantum} have explored the quantum origin of this nonlinear response by introducing an intrinsic effect of the dipole moment of the Berry curvature. This Berry curvature dipole (BCD) can be observed in time-reversal invariant systems, but its non-zero value is strictly protected by the breaking of inversion symmetry of the crystal. Moreover, symmetry-based indicators are also crucial in determining the strength of BCD in a noncentrosymmetric system. For example, a uniaxial strain reduces the symmetry of TMDs and gives rise to enhanced BCD \cite{you2018berry,zhou2020highly,son2019strain}. Similar enhancements are also observed for few-layer TMDs \cite{ma2019observation,kang2019nonlinear,huang2020giant,joseph2021topological}, where the lowering of symmetry is the result of the stacking of monolayers. It has further been observed that the electric field can efficiently tune the anomalous NHE in low-symmetry TMDs \cite{zhang2018electrically}. Moreover, a pressure-driven topological phase transition in three-dimensional bismuth tellurium iodine (BiTeI) with a strong Rashba effect is also predicted to lead to a large BCD \cite{facio2018strongly}. In principle, the BCD-induced NHE can be considered as a second-order response to the electric field in the system's plane. The combined effect of this in-plane electric field and BCD is responsible for several exotic physical properties, such as giant magneto-optical effects \cite{liu2020anomalous}, orbital valley magnetization \cite{son2019strain}, non-linear Nernst effects \cite{yu2019topological,zeng2019nonlinear}, and thermal Hall effect \cite{zeng2020fundamental}.

In general, large Berry curvature segregation occurs at the Brillouin zone (BZ) points, where two bands nearly touch each other. The shape of the Bloch states rapidly modifies near such narrow-gap points of the BZ. Therefore, massive tilted Dirac cones  \cite{sodemann2015quantum,nandy2019symmetry,du2019disorder} or Weyl cones \cite{kumar2021room,zeng2021nonlinear,matsyshyn2019nonlinear,singh2020engineering,roy2021non} are the natural choices for realizing sizeable BCD. In these systems, the BCD and corresponding NHE systematically provide the geometrical information of Bloch wavefunctions even under TRS. Moreover, Battilomo \textit{et al.} have revealed that the Fermi surface warping triggers appreciable BCD in uniaxially strained monolayer and bilayer graphene \cite{battilomo2019berry}. It is worth mentioning that the magnitude of BCD in the warped graphene systems is comparable with that of the TMDs. Further, the merging of Dirac points near the Fermi level can also lead to non-zero BCD even in the absence of any tilt or warping term \cite{samal2021nonlinear}. The BCD and the unconventional NHE drive the understanding of topological physics and quantum transport phenomena to the nonlinear domain. This generalization opens up many exciting prospects for direct applications, such as nonlinear photocurrents \cite{xu2018electrically} and terahertz radiation detection \cite{zhang2021terahertz}. 

In a pioneering work, Kane and Mele \cite{kane2005quantum} first explored the fascinating quantum spin Hall effect (QSHE) in graphene. However, in reality, the QSHE in graphene is not experimentally accessible because of the negligible strength of spin-orbit coupling (SOC). The strength of SOC largely determines the occurrence of helical edge states with well-defined spin texture in topological insulators. In real systems, the requirement of large SOC is considerably fulfilled by the experimental realization of `graphene counterparts' -- silicene, germanene, and stanene \cite{balendhran2015elemental,vogt2012silicene,derivaz2015continuous,zhu2015epitaxial}. Silicene, germanene, and stanene are two-dimensional (2D) Dirac materials having a buckled honeycomb geometry. The buckling can be exploited by employing a transverse electric field that can tune the electronic band structure, particularly, the band gap \cite{mak2009observation}. Furthermore, an electric field driven topological phase transition from QSH to a normal insulating (NI) state is a primary characteristic of these systems \cite{ezawa2015monolayer,deng2018epitaxial,qin2021strain}. 

In this work, we discover a large and electrically switchable NHE in these elemental buckled honeycomb lattices silicene, germanene, and stanene. Using tight-binding calculations, in conjunction with symmetry arguments, we explore the tunability of the BCD in these systems, particularly near the topological phase transitions. We demonstrate a giant enhancement of the BCD near the electric field tuned topological critical points and connect it to the underlying variations of the Berry curvature. Our findings put forward a new class of systems to explore nonlinear topological phenomena, and highlight an as-yet-unexplored aspect of elemental buckled honeycomb lattices. We hope that our work motivates experimental as well theoretical work along this front in the near future. 

\section{Methodology}

\begin{figure*}
\centering
\includegraphics[scale=0.22]{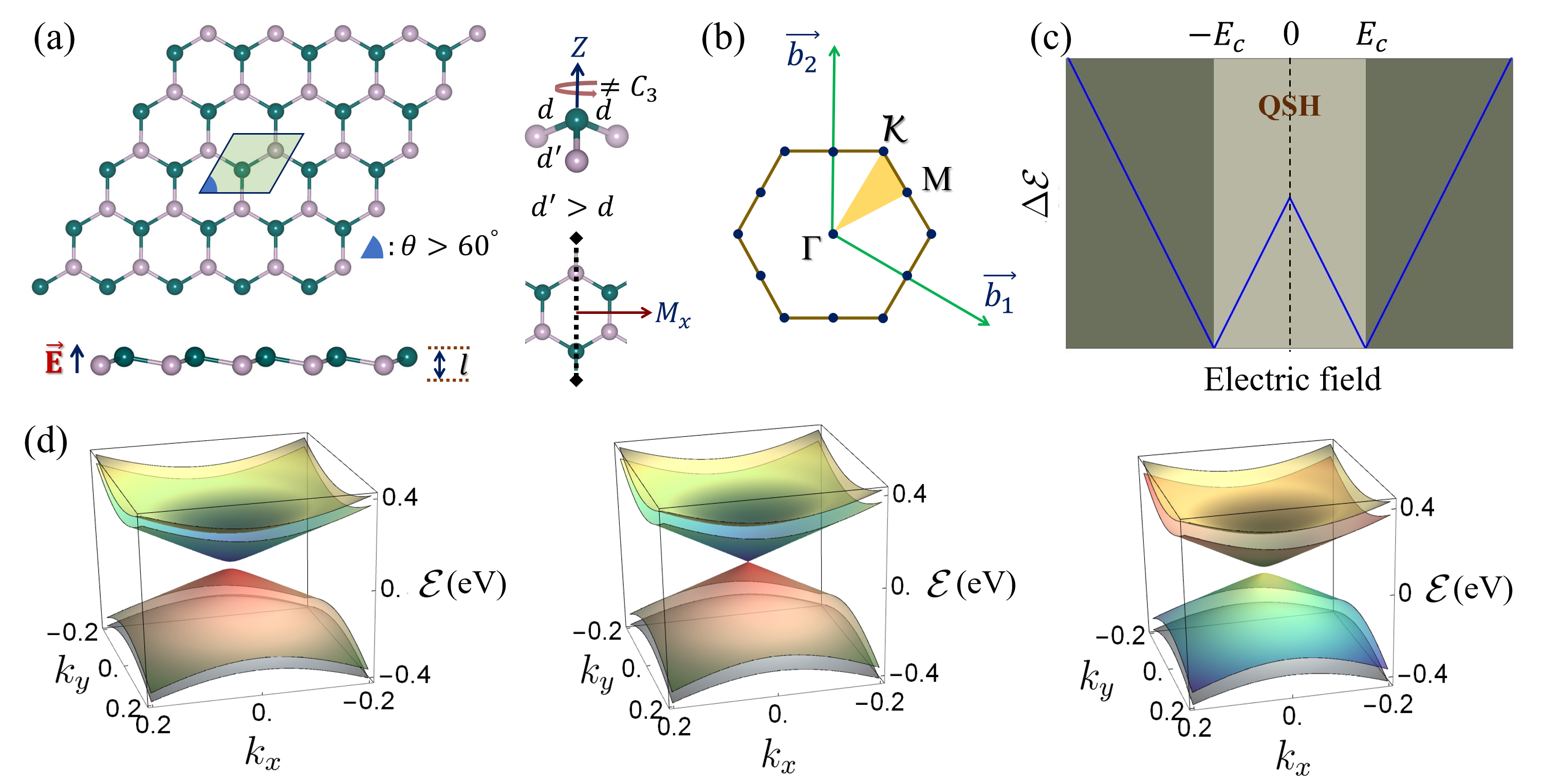}
\caption{\textbf{Buckled honeycomb lattices and their topological nature.} (a) The top and side view of buckled strained honeycomb structures -- silicene, germanene, and stanene. One of the three nearest bonds is stretched (up to $\approx 2\%$). Therefore, the angle between the translation vectors increases from the usual 60$^{\mathrm{o}}$. A transverse electric field ($\vec{E}$) adequately exploits the buckling and breaks the inversion symmetry of the lattice. Different colors indicate two sub-lattices for clarity. The elongated bond destroys the $C_3$ rotational symmetry but protects the mirror axis ($M_x$) in the direction perpendicular to the stretching. (b) The Brillouin zone of the symmetry reduced buckled honeycomb structures. The coordinates of the symmetry point $\mathcal{K}$ are distinct from that of the original systems. (c) The phase diagram of the perturbed system ($\approx 2\%$ stretch) resembles the pristine structure. The blue line represents the variation of band gap with external electric field. Appreciable spin-orbit interaction gives rise to the quantum spin Hall (QSH) phase with a finite band gap ($\Delta \mathcal{E}$). The band gap decreases with increasing electric field, and above a critical value ($ \lvert E_c \rvert$), a topological phase transition from topological insulator to normal insulator (NI) occurs. The bottom panel (d) Band structures at different electric fields. Two spin bands touch the Fermi level for the electric field strength $ \lvert E_c \rvert$. At this point, the system behaves as a semimetal with gapless bands. The systems exhibit a band gap below and above the critical field. The spin degeneracy of the bands is broken due to the presence of a finite electric field.} \label{fig:1}
\end{figure*}

The nonlinear current in response to an oscillating electric field $\vec{E}(t) = Re\{\vec{E}_0 e^{i \omega t}\}$, with a magnitude $E_0$ and frequency $\omega$ can be expressed as $J_{\alpha} = Re\{J_{\alpha}^{(0)} + J_{\alpha}^{(2\omega)} e^{i 2 \omega t}$\} \cite{sodemann2015quantum}. Therefore, the response current has been clearly decomposed into a static part (rectified current) $J_{\alpha}^{(0)} = \chi_{\alpha \beta \gamma} \vec{E}_{\beta} \vec{E}_{\gamma}^{*}$ and a double frequency oscillating part (second harmonic) $J_{\alpha}^{(2 \omega)} = \chi_{\alpha \beta \gamma} \vec{E}_{\beta} \vec{E}_{\gamma}$. Under time-reversal symmetric conditions, the nonlinear conductivity tensor ($\chi_{\alpha \beta \gamma}$) depends on the momentum derivative of the Berry curvature over the occupied states as follows

\begin{equation}
    \chi_{\alpha \beta \gamma} = - \epsilon_{\alpha \delta \gamma} \frac{e^3 \tau}{2(1+i\omega \tau)} \int_{k} [d\vec{k}] f_0 \left(\frac{\partial \Omega_\delta}{\partial k_\beta}\right).
    \label{eq:chi}
\end{equation}

Here, $e$, $\tau$, $\epsilon_{\alpha \delta \gamma}$, $f_0$ and $\Omega_\delta$ represent the electron charge, scattering time, Levi-Civita symbol, equilibrium Fermi-Dirac distribution, and Berry curvature component along $\delta$, respectively with $\alpha, \beta, \gamma, \delta \in \{x,y,z\}$. Here the integration is performed with respect to $[d\vec{k}]$, which has the expression $d^{d} k / (2 \pi)^{d}$ in $d$ dimensions. BCD can be defined in reciprocal space as $D_{\alpha \beta} = \int_{k} [d\vec{k}] f_0 \left(\partial \Omega_\beta / \partial k_\alpha \right)$. Particularly, in 2D materials, only the out-of-plane ($z$) component of Berry curvature is non-vanishing, i.e., $\Omega_\beta \equiv \Omega_z$. In the framework of the well-known Kubo formalism, this $\Omega_z (\vec{k)}$ has the following form \cite{xiao2010berry}

\begin{equation}
    \Omega_z (\vec{k}) = 2 i \sum_{i \neq j} \frac{\langle i|{\partial\hat{H}/\partial k_x}|j\rangle \langle j|{\partial\hat{H}/\partial k_y}|i\rangle}{\left(\varepsilon_i - \varepsilon_j\right)^{2}},
\label{eq:bc}
\end{equation}

where $ \varepsilon_i$ and $\varepsilon_j$ are the eigenenergies of the Hamiltonian $\hat{H}$ with eigenstates $|i\rangle$ and $|j\rangle$, respectively. The methodology discussed above for calculating BCD is implemented in the Wannier-Berri package \cite{tsirkin2021high}, which is compatible with the PythTB module \cite{Pythtb}. It is worth mentioning that the Berry curvature of time-reversal invariant systems is an odd function of momentum, i.e., $\mathcal{T}^{\dagger} \Omega_z(-\vec{k}) \mathcal{T} = - \Omega_z(\vec{k})$, where $\mathcal{T}$ is the time-reversal operator. In contrast, BCD is even under the above situation, as it satisfies $\mathcal{T}^{\dagger} D_{\alpha \beta}(-\vec{k}) \mathcal{T} = D_{\alpha \beta}(\vec{k})$. Therefore, it is clear that BCD can manifest a significant anomalous electronic response, NHE, even in the presence of TRS.

In order to support the tight-binding results, first-principles calculations are carried out based on the density functional theory (DFT) framework as implemented in the {\sc quantum espresso} code \cite{QE-2017,QE-2009}. A kinetic energy cut-off of $40$ Ry is considered, using the ultrasoft pseudopotentials \cite{PhysRevB.41.7892} to describe the core electrons, including spin-orbit coupling interactions. We used the Perdew-Burke-Ernzerhof (PBE) form for the exchange-correlation functional \cite{perdew1996generalized}. The Brillouin zone is sampled over a uniform $\Gamma$-centered $k$-mesh of $8\times8\times1$, and the monolayers were modeled with a 15 \AA~vacuum along the $z$-direction to avoid any spurious interaction between the periodic images. To study the topological properties, maximally localized Wannier functions (MLWFs) were computed to derive a tight-binding model from the \textit{ab-initio} calculations, with complete $s,p$ orbitals as the basis, using the {\sc wannier90} code \cite{mostofi2014updated}. Further calculation of $\mathbb{Z}_2$ topological invariants and analysis of the edge spectra is performed using the WannierTools code \cite{WU2017}.

\begin{figure*}
\centering
\includegraphics[scale=0.7]{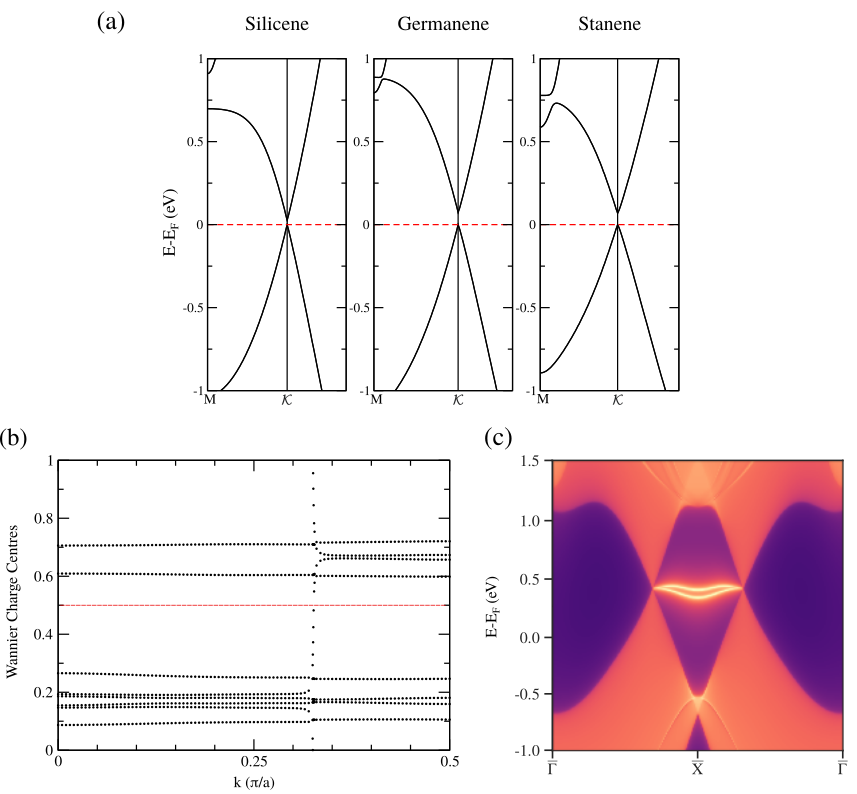}
\caption{\textbf{The electronic properties obtained by density functional theory methods.} (a) The band structures of 2\% strained buckled honeycomb lattices exhibit a band splitting at $\mathcal{K}$ point. (b) Wannier charge centers for strained silicene demonstrate the nontrivial $\mathbb{Z}{_2}$ index of the system. (c) The protected edge states have been calculated in the nontrivial topological region of the phase space.} 
\label{fig:dft2}
\end{figure*}

\section{Results and discussion}

\subsection{Model and symmetry analysis} Next, we introduce the tight-binding (TB) model Hamiltonian for buckled honeycomb lattices used in this work. In spite of the similarity in their basic geometry, the buckled honeycomb lattices primarily differ in their bond lengths ($d$), hopping integrals ($t$), and lattice parameters ($a = \sqrt{3} d$). Further, the strength of SOC ($\lambda_{SO}$) and buckling height ($l$) between the two sublattices also depend on the atomic number of the constituent atom and vary from system to system. All the above mentioned parameters of buckled honeycomb lattices are compared with that of graphene in Table \ref{tab:tabls}. In our study, we shall extensively use these values in Table \ref{tab:tabls} to extract the system specific information, while also using general symmetry arguments. In the presence of buckling, the transverse electric field ($E_{z}$) assigns different mass terms to the two sublattices. This difference in mass terms, in turn, breaks the inversion symmetry of the system. The generalized TB Hamiltonian \cite{ezawa2015monolayer} of buckled honeycomb lattices in the presence of an external transverse electric field can be expressed as

\begin{align}
     \hat{H} = & - t \sum_{\langle ij\rangle,\sigma} c_{i,\sigma}^{\dagger}c_{j,\sigma} + i \frac{\lambda_{SO}}{3\sqrt{3}} \sum_{\langle\langle ij\rangle\rangle ,\sigma} \sigma \zeta_{ij} c_{i,\sigma}^{\dagger}c_{j,\sigma} \nonumber \\
     & -l \sum_{i,\sigma} \nu_{i} E_{z} c_{i,\sigma}^{\dagger}c_{i,\sigma}.
\label{eq:ham}
\end{align}

Here $\langle ij\rangle$ and $\langle\langle ij\rangle\rangle$ indicate hopping between $i$ and $j$ sites up to nearest and next-nearest neighbours. Further, $\sigma$ represents spin degrees of freedom and denotes either $\uparrow$ ($+1$), or $\downarrow$ ($-1$) spin. The SOC and staggered sublattice potential, $\zeta_{ij}$ $(= \pm 1$) and $\nu$ $(= \pm 1$) explicitly depend on the direction (clockwise or anticlockwise) of hopping and type of sublattice, respectively. Furthermore, Eq.~\ref{eq:ham} reveals that buckled honeycomb lattices exhibit a topologically nontrivial band gap, $\Delta \mathcal{E} = 2 \lambda_{SO}$, in the absence of an external electric field \cite{ezawa2015monolayer}. Application of the electric field reduces this value of the band gap to zero at a topological critical point $E_C = \lambda_{SO}/l$, where the system behaves like a semimetal. Beyond this point, $\Delta \mathcal{E}$ again increases and gives rise to a topologically trivial phase.
The tight-binding parameters for Eq.~\ref{eq:ham} vary across different systems we considered. The parameters for silicene, germanene, and stanene are compared with graphene in Table \ref{tab:tabls}. Because of the negligible spin-orbit coupling and planar geometry, graphene does not show an electrically tunable quantum spin Hall state.

\begin{table}
    \centering
    \caption{The structural and tight binding model parameters of silicene, germanene, and stanene are compared to that of graphene. Here, $a$, $t$, $\lambda_{SO}$, and $l$ represent the lattice parameter, hopping integral, SOC strength, and buckling heights, respectively. The values are taken from Ref. \cite{liu2011low}.}
    \setlength{\tabcolsep}{0.5em}
    \def\arraystretch{1.2}%
    \begin{tabular}{lllll}
    \hline
    \hline
     Systems & $a$ (\AA{}) & $t$ (eV)  & $\lambda_{SO}$ (meV) & $l$ (\AA{}) \\
     \hline
     \hline
     Graphene & 2.46 & 2.8 & 10$^{-3}$ & 0.00 \\
     Silicene & 3.86 & 1.6 & 3.9 & 0.23 \\
     Germanene & 4.02 & 1.3 & 43.0 & 0.33 \\
     Stanene & 4.70 & 1.3 & 100.0 & 0.40 \\
     \hline
     \end{tabular}
     \label{tab:tabls}
\end{table}

To trace this topological transition, we initiate our calculations in the presence of a non-zero electric field. The electric field helps us to attain the essential criterion of inversion symmetry breaking for BCD. Regardless of the breaking of inversion symmetry in buckled honeycomb lattices, we obtain a vanishing value of BCD at different electric field strengths. This can be understood from the crystallographic symmetry of the buckled honeycomb lattices. The buckling in buckled honeycomb lattices primarily eliminates the $C_{6}$, $\sigma_{h}$, and three $\sigma_{d}$ symmetry elements of the planar honeycomb lattices ($D_{6h}$). Consequently, the system possesses only three $C_{2}$ rotational axes perpendicular to the principal axis of symmetry. Further, three mirror planes of the buckled system bisect the angle between each neighboring pair of these $C_{2}$ rotational axes. Therefore, the groups of wavevectors in buckled honeycomb lattices are $D_{3d}$ (point group of buckled honeycomb lattices) and $D_{3}$ at symmetry points $\Gamma$ and $K$ respectively. The order of the finite group $D_{3}$ is $3 \times 2 = 6$ with 3 rotational and 3 reflection symmetry elements. In particular, the rotational symmetry elements are $2 \pi$, $2 \pi/3$, and $4 \pi/3$ rotation about the $C_{3}$ axis, while the reflection symmetry elements represent the symmetry planes ($\sigma_{d}$) passing through the 3 medians of the equilateral triangle. The character tables are presented in Table S1 and Table S2 of ESI~\cite{Suppl_mat}. The presence of two or more mirror axes in two-dimensional buckled honeycomb lattices (here, three) relates non-linear Hall conductance to a null pseudovector field \cite{ortix2021nonlinear}, so that the BCD is zero. Fundamentally, the maximum permitted symmetry in two dimensions for the occurrence of non-vanishing BCD is a single mirror line. This symmetry analysis motivates us to reduce the symmetry of buckled honeycomb lattices down to a single mirror axis by minimal operations. We have achieved this by applying a small uniaxial strain that elongates one nearest neighbour bond ($d^{\prime}$) compared to the other two ($d$) as depicted in Fig~\ref{fig:1}(a). As a representative value we have chosen to $d^{\prime}$ to be different from $d$ by 2\%. Application of such a strain does not break the inversion symmetry of these systems but essentially reduces its rotational symmetry. In the above case, $C_{3}$ rotational symmetry is destroyed along with two $\sigma_{d}$ reflection symmetry planes. The strained buckled honeycomb lattices possess only one $\sigma_{d}$ symmetry element, which corresponds to $M_{x}$ mirror symmetry plane in our case. We expect that these strained buckled honeycomb lattices will be suitable candidates for obtaining sizable BCD in the presence of an external transverse electric field.

\begin{figure}
\centering
\includegraphics[scale=0.18]{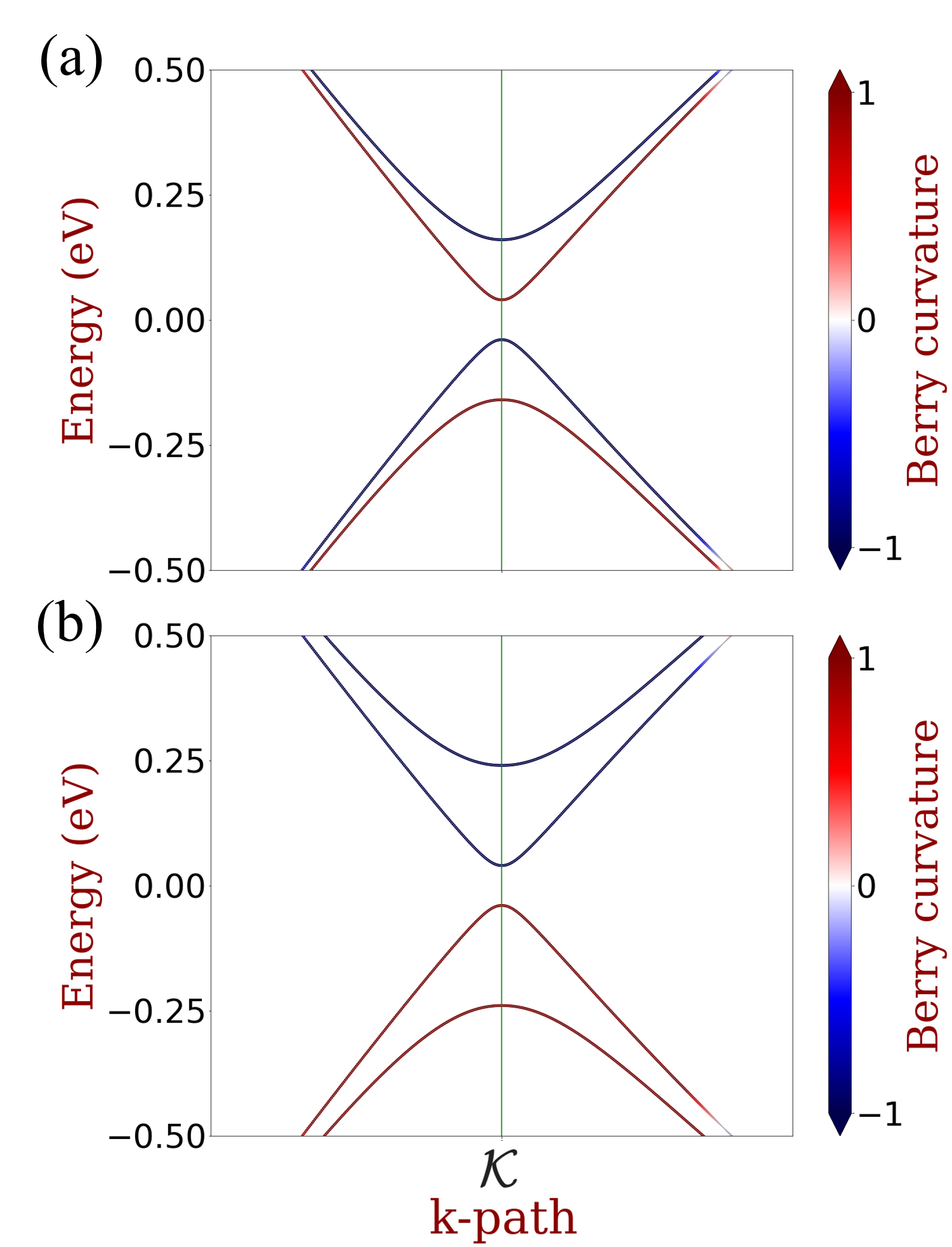}
\caption{\textbf{Berry curvature and electrical field.} The Berry curvatures of the four bands near $\mathcal{K}$ point are shown for stanene under (a) 0.15 V/\AA{} (non-trivial phase) and (b) 0.35 V/\AA{} (trivial phase) electric field strength. The spin degeneracy of the bands is lifted under the electric field. A clear flipping of the Berry curvature of two bands indicates the topological phase transition. The color scale is shown on the right.} 
\label{fig:2}
\end{figure}

\begin{figure}[H]
\centering
\includegraphics[scale=0.45]{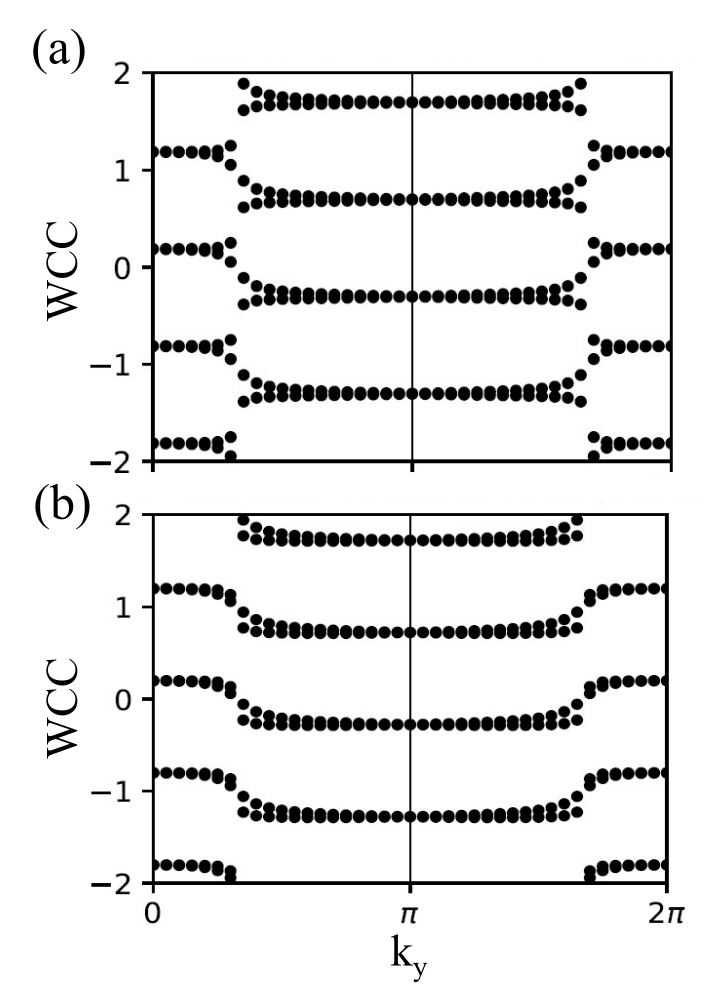}
\caption{\textbf{Wannier charge centers.} The Wannier charge centers (WCCs) are plotted for stanene under (a) 0.15 V/\AA{} (non-trivial phase) and (b) 0.35 V/\AA{} (trivial phase) electric field strengths. In the first case, a single line parallel to $k_y$ crosses an odd number of charge centers in the half of the BZ, giving rise to odd $\mathbb{Z}_2$ invariant. This odd $\mathbb{Z}_2$ confirms the non-trivial topological state of the material. (b) In the second case, due to an even number of crossing points, the WCC reveals that the $\mathbb{Z}_2$ invariant is now even, and the phase is trivial.} 
\label{fig:3}
\end{figure}

The unit cell of the strained buckled honeycomb lattices is defined by a new set of lattice vectors: $\vec{a}_1= \sqrt{3}d \hat{x}$ and $\vec{a}_2=\sqrt{3}d /2 \hat{x} + (3 d /2 + 2d/100) \hat{y}$. As a result, the angle, $\theta$, between lattice parameters slightly increases than the usual $60^{\circ}$ of the pristine case. In the reciprocal space, the strain changes the high symmetry point $K$ $(0.3333,0.6667)$ to an equivalent point $\mathcal{K}$ $(0.3404,0.6702)$ of the BZ as shown in Fig.~\ref{fig:1}(b). Moreover, the band minimum or maximum of strained buckled honeycomb lattices occur at this new point $\mathcal{K}$, near which the band dispersion is linear. Similar to the unperturbed case, the band gap of the slightly strained buckled honeycomb lattices is $2\lambda_{SO}$ in the absence of an electric field. Furthermore, the applied electric field controls the band gap as $\Delta \mathcal{E} \approx 2 l |E_{z} - \lambda_{SO}/l|$ as indicated in Fig.~\ref{fig:1}(c). Note that the strained systems are semimetallic ($\Delta \mathcal{E}$ is zero) at a critical electric field $E_C \approx \lambda_{SO}/l$. The low-energy effective Hamiltonian $\hat{H}_{\mathrm{eff}}$ around $\mathcal{K}$ can be written as follows

\begin{equation}
    \hat{H}_{\mathrm{eff}} \approx  \begin{bmatrix}
\mu + h_{11} & v_x k_{x}+i v_y k_{y} \\
v_x k_{x} - i v_y k_{y} & \mu - h_{11} 
\end{bmatrix},
\label{eq:eff}
\end{equation}

where $v_x = v_y = v_{F}$ is the Fermi velocity. The matrix elements $h_{11}$ and $\mu$ are related to the Pauli spin matrix $\sigma_z$ by the relations $h_{11} =  - \lambda_{SO}\sigma_{z}$ and $\mu = l E_{z} \sigma_{z}$. The three-dimensional band structures obtained using $\hat{H}_{\mathrm{eff}}$ around $\mathcal{K}$ are illustrated in Fig.~\ref{fig:1}(d). As expected, the band structures of strained buckled honeycomb lattices exhibit band opening on both sides of the critical field $E_{C}$ similar to the pristine case (see also, Fig. S1 of ESI~\cite{Suppl_mat}).

The DFT band structures for all the three strained systems, given in Fig.~\ref{fig:dft2}(a), also indicate that the strained buckled honeycomb lattices invariably exhibit a band gap at the $\mathcal{K}$ point. Further, the nontrivial topological nature of the band gap has been confirmed by the Wannier charge centers plots ($\mathbb{Z}{_2}=1$) and symmetry protected edge states given in Fig.~\ref{fig:dft2}(b) and Fig.~\ref{fig:dft2}(c), respectively. These results are in excellent agreement with our tight-binding model findings.

\subsection{Berry curvature} We then calculated the Berry curvature using Eq.~\ref{eq:bc}, to understand the topological aspects of the above-mentioned gapped states of our strained buckled honeycomb lattices. It has been observed that the sign of Berry curvature of two spin states near the Fermi level, i.e., conduction band (CB) and valence band (VB) flips when electric field strength crosses the critical value $E_C$ \cite{zhang2018electrically}. A clear example of this flipping of Berry curvature for strained stanene with $E_C \approx 0.2499$ V/\AA{} is shown in Fig.~\ref{fig:2}. It is evident from Fig.~\ref{fig:2}(a) and (b) that the Berry curvature distribution of the CB and VB are distinct for the electric field strength 0.15 V/\AA{} and 0.35 V/\AA{}. In particular, the Berry curvature of the VB (CB) is negative (positive) under 0.15 V/\AA{} electric field, while the same is positive (negative) for the field strength 0.35 V/\AA{}. In a similar vein, we have also observed the Berry curvature exchange between the VB and CB for strained silicene ($E_C \approx 0.0170$ V/\AA{}) and strained germanene ($E_C \approx 0.1303$ V/\AA{}). Note that the Berry curvature of all the systems tends to diverge while reaching the critical point from both sides. All the results mentioned above are certainly the signatures of a topological phase transition of the buckled buckled honeycomb lattices. 

To confirm the presence of distinct topological phases in strained buckled honeycomb lattices, the $\mathbb{Z}_2$ invariant has been evaluated from the Wannier charge centers (WCC) \cite{yu2011equivalent,soluyanov2011computing}. Fig.~\ref{fig:3}(a) indicates that a straight line parallel to $k_y$ intersects WCC an odd number of times ($\mathbb{Z}_2=1$) in half of the BZ of strained stanene under 0.15 V/\AA{} electric field. On the other hand, the same has an even number of intersections ($\mathbb{Z}_2 =0$) for the case of 0.35 V/\AA{} electric field [Fig.~\ref{fig:3}(b)]. The $\mathbb{Z}_2$ invariant establishes that the external electric field drives the strained stanene from a non-trivial topological state to a trivial one. Similar results have also been observed for strained silicene and germanene structures. Therefore, the electric field dependent topological phase transition in BHLs is robust to this applied strain. On that account, the required criteria for tracing the topological phase transition in BHLs by BCD are entirely fulfilled.

\begin{figure}[htbp]
\centering
\includegraphics[scale=0.24]{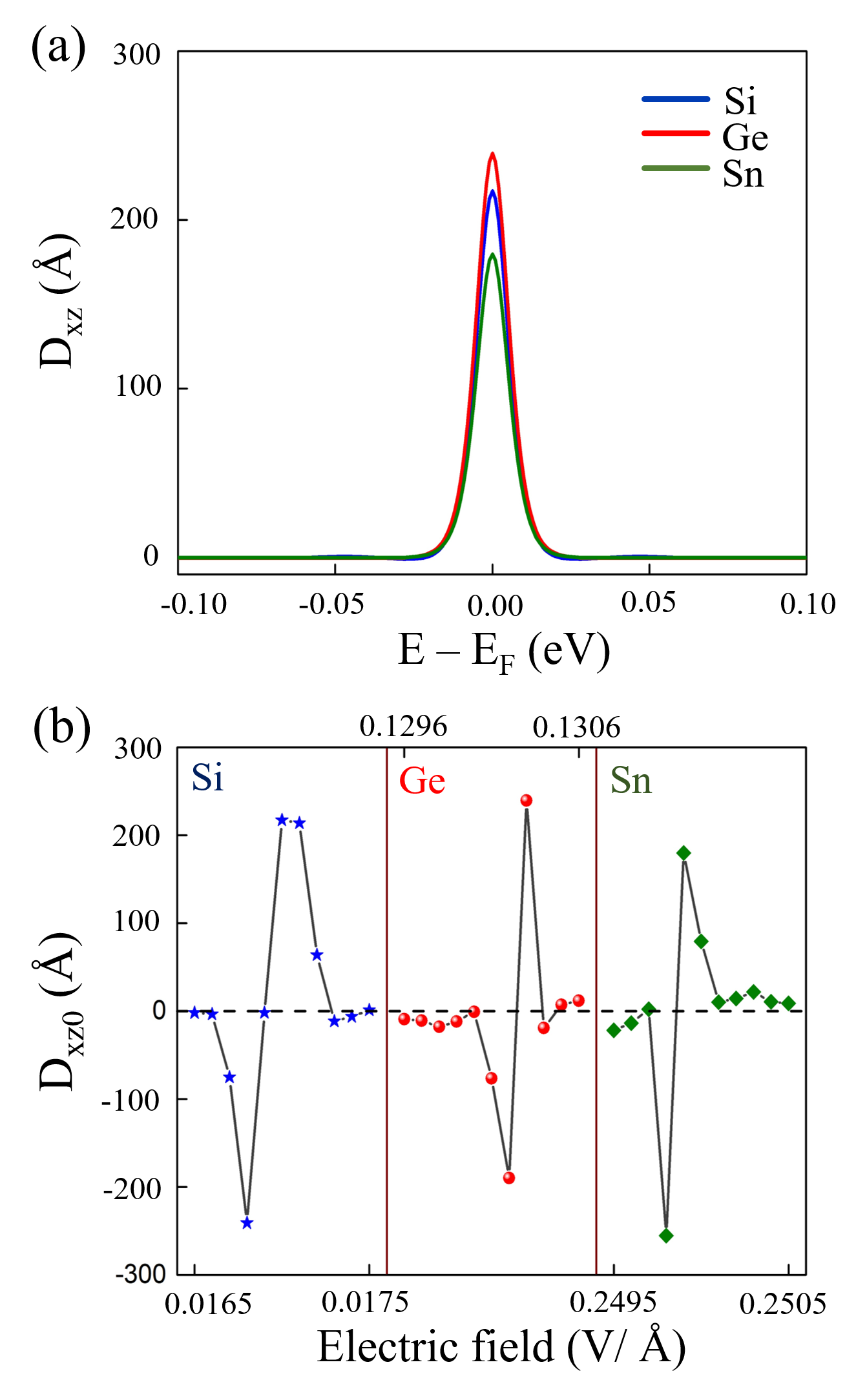}
\caption{\textbf{BCD and its variation with electric field.} (a) The variation of BCD with the energy exhibits a peak near the Fermi level for all the systems. Here, BCD for the silicene under 0.0170 V/\AA{} electric field, germanene under 0.1303 V/\AA{} electric field, and stanene under 0.2499 V/\AA{} electric field are given as an example. Only the $D_{xz}$ component of the BCD tensor is non-zero from symmetry-based indicators. (b) The variation of BCD at the Fermi level ($D_{xz0}$), with the electric field for silicene, germanene, and stanene, shows a peak around the corresponding topological critical point. The sign of BCD also changes on either side of the topological transition.} 
\label{fig:4}
\end{figure}

\subsection{Berry curvature dipole} This rapid change with electric field in sign and magnitude of Berry curvature in momentum space strongly suggests the possibility of large and tunable BCD in strained buckled honeycomb lattices. We next explore this aspect in these systems by calculating the BCD in the presence of an external electric field. As we noted, the strained buckled honeycomb lattices possess only one mirror symmetry $M_x$. In the presence of $M_{x}$ symmetry, $k_x$ and $k_y$ transform as odd and even parameters, respectively. $\Omega_{z}$ is the only non-zero component of Berry curvature in these 2D systems, which is odd in momentum. The above observations immediately show that the gradient of $\Omega_z$ along $k_x$, or, $d_{xz}=\partial \Omega_z / \partial k_x$ will be the only even term. In other words, the momentum integrated $d_{xz}$, i.e., $D_{xz}$ will be the only non-zero BCD tensor component in strained buckled honeycomb lattices. We find that all these strained buckled honeycomb lattices exhibit a substantial BCD in the presence of electric field induced inversion symmetry breaking. Moreover, we have discovered a generic feature of $D_{xz}$ in these systems -- the contribution of $D_{xz}$ is enormous near the Fermi level for all the strained buckled honeycomb lattices as shown in Fig.~\ref{fig:4}(a). This giant BCD value can be well-explained by the high concentration of Berry curvature near the Fermi level. Furthermore, the sign of BCD is found to be reversed when the direction of electric field, $E_z$, is flipped. A Fermi smearing of 40 K is considered for quantitative discussions. The band gap of these systems gradually decreases to zero at $E_c$ and then again increases linearly. This band gap variation results in a maximum BCD response near the topological transition point, as we found. In particular, the maximum BCD of strained germanene at the Fermi level is $D_{xz0} = 239.35$ \AA{}, which is larger than that of silicene ($D_{xz0} = 217.16$ \AA{}). In contrast, the maximum BCD of stanene $D_{xz0}=179.73$ \AA{} does not follow the above trend of with increasing SOC. Similar results have been reported for Weyl semimetals TaAs, TaP, NbAs, and NbP \cite{zhang2018berry}. In comparison, the maximum value of BCD for other 2D materials, such as strained monolayer graphene \cite{battilomo2019berry}, bilayer graphene \cite{battilomo2019berry}, WTe$_2$ \cite{zhang2018electrically}, and MoTe$_2$ \cite{zhang2018electrically} are approximately 0.01 \AA{}, 10 \AA{}, 60 \AA{}, and 80 \AA{}, respectively. Therefore, strained buckled honeycomb lattices can provide an intriguing platform to achieve large and tunable BCD. In particular, dual-gated, encapsulated devices can be fabricated based on the strained ($\sim$ 2\%) buckled honeycomb lattices with controllable chemical potential and transverse electric field setup for realizing BCD at low temperature ($\sim$ 100 K). The alternating electric field applied to the device will result in a non-linear voltage, with doubled frequency, that can be measured using a sensitive lock-in amplifier.

The diverging nature of BCD at topological transition point can be well understood from the following discussion. It is evident from Eq.~\ref{eq:eff} that spin-orbit coupling gives rise to massive Dirac cones by opening a gap $\Delta \mathcal{E} \equiv \Delta \approx 2 l |E_{z} - \lambda_{SO}/l|$ in the energy spectra. In the case of these strained buckled honeycomb lattices, another term proportional $k_y$ ($\lambda k_y$) can be included to address the non-isotropic dispersion arising from the strain. The term proportional to $\lambda$ gives the anisotropic velocity, depending on the applied strain. It is worth noting that this $\lambda$ term is responsible for the non-zero value of the Berry curvature dipole. Further, two non-equivalent massive Dirac cones are related by the time-reversal symmetry. Hence, only the out-of-plane component of Berry curvature is non-zero, which is segregated in these two valleys with a different sign. Furthermore, the small external strain is responsible for a perturbed Berry curvature distribution and assigns different weights to it in the Brillouin zone. We can write a low energy model Hamiltonian for the system considering the states near the Fermi level as

\begin{equation}
   H = \left( v_x k_x \sigma_y - \tau v_y k_y \sigma_x \right) + \Delta \sigma_z + \tau \lambda k_y.
    \label{eq:low}
\end{equation}

Here valley index $\tau = \pm$, $\sigma_x$ and $\sigma_y$ are Pauli spin matrices. The Hamiltonian given above allows only the $M_x$ crystal symmetry, where $k_x \rightarrow - k_x$ symmetry is preserved. The dispersion relation obtained using Eq.~\ref{eq:low} is obtained as

\begin{equation}
 \varepsilon_{\pm}(k_x,k_y)= \tau \lambda k_y \pm \left( \Delta^2 + v_{x}^2 k_{x}^{2}+ v_{y}^2 k_{y}^{2} \right)^{1/2}.
    \label{eq:dis}
\end{equation}

Now, the Berry curvature $\Omega_z (\vec{k})$ can be calculated using Eq.~\ref{eq:bc}. For buckled honeycomb lattices $\Omega_z (\vec{k})$ can be expressed as

\begin{equation}
 \Omega_z (k_x,k_y) = \pm \frac{1}{2} \frac{\tau v_x v_y \Delta}{\left( \Delta^2 + v_{x}^2 k_{x}^{2}+ v_{y}^2 k_{y}^{2} \right)^{3/2}}.
\label{eq:bc1}
\end{equation}

The BCD is related to the Berry curvature by $D_{\alpha \beta} = \int_{k} [d\vec{k}] f_0 \left(\partial \Omega_\beta / \partial k_\alpha \right)$. It is then straightforward to write down the expression of Berry curvature dipole mentioned above in terms of Berry curvature dipole density $d_{\alpha \beta}$ as follows

\begin{equation}
D_{\alpha \beta} = - \sum_{n} \int \frac{\partial f_n (\vec{k})}{\partial \varepsilon_n(\vec{k})} v_\alpha \Omega_{n\beta} (\vec{k}) d[\vec{k}] = \int  \frac{\partial f_n (\vec{k})}{\partial \varepsilon_n(\vec{k})} d_{\alpha \beta} d[\vec{k}].
   \label{eq:bcddensity}
   \end{equation}

In the above, $f_n (\vec{k})$ is the distribution function and the velocity $v_\alpha$ can be obtained by $\partial \varepsilon_{n}(\vec{k})/ \partial k_{\alpha}$. The BCD density then has the following expression

\begin{equation}
 d_{\alpha \beta} = - \sum_{n} v_\alpha \Omega_{n\beta} (\vec{k}).
\end{equation}

The partial differentiation gives the delta function in the low energy limit, which indicates that Berry curvature dipole is a Fermi surface property. We note that the band gap of the system is $E_g= 2 \left( \Delta^2 + v_{x}^2 k_{x}^{2}+ v_{y}^2 k_{y}^{2} \right)^{1/2}$. From the above discussion it is evident that

\begin{equation}
\lim_{E_g \to 0} d_{xz} = \frac{1}{2} \lim_{E_g \to 0} \sum_{n} \frac{  v_\alpha \tau v_x v_y \Delta}{\left( \Delta^2 + v_{x}^2 k_{x}^{2}+ v_{y}^2 k_{y}^{2} \right)^{3/2}} \rightarrow \infty.    
\label{eq:infty}
\end{equation}

Therefore, we find that the BCD density predominantly varies as $\sim$ $1/\Delta^2$ and diverges near the topological critical point where the band gap closes.

\subsection{Nonlinear thermal Hall effect}
Similar to the electrical Hall effect, the thermal Hall current also vanishes under time reversal symmetric condition in the linear response regime. However, in case of the strained honeycomb lattices the perturbed distribution function gives rise to nonlinear thermal Hall effect. The thermal Hall current, $j_{a}^{T}$, in the nonlinear regime can be obtained using Boltzmann equation as given below  

\begin{equation}
j_{a}^{T} = - \kappa_{abd} \Delta_b T \Delta_d T.
\label{eq:nth}
\end{equation}

Here $\Delta T$ represents temperature gradient and the subscripts \{$a, b, d$\} $\in$ \{$x, y, z$\} stand for the components. The coefficient of nonlinear thermal Hall effect is denoted by $\kappa_{abd}$. From the symmetry analysis it is clear that the temperature gradient along a direction normal to $M_x$ gives rise to nonlinear thermal Hall effect in the perpendicular direction. The intrinsic contribution of the nonlinear thermal Hall coefficient, $\kappa^{BCD}_{abd}$, comes from the BCD \cite{zeng2020fundamental} given as follows

\begin{equation}
\kappa^{BCD}_{abd} = \zeta T^2 G^{1}_{0}(\mu) + \mathcal{O}[T^4],
\label{eq:nthk}
\end{equation}

where $\zeta = \frac{7 \tau_0 \pi^4 k_{B}^{4}}{14 \hbar^2}$ is a constant under the constant relaxation time ($\tau_0$) approximation. The chemical potential dependent parameter $G^{1}_{0}(\mu) = \partial G_{0}(\mu)/ \partial \mu$ is related to the BCD by

\begin{equation}
G_{\alpha \beta}(\mu) = \int [d k] \delta\left(\varepsilon - \varepsilon (k)\right) \Omega_{\alpha}(k) 
\frac{\partial \varepsilon (k)}{\partial k_\beta}.
\label{eq:nthkf}
\end{equation}

In presence of disorder, relaxation time $\tau_0$ can be written as $\tau_0^{-1} = n V_{0}^{2} (\mu^2 + 3 \Delta^2)/4\hbar \mu v_x v_y $, where, $n$ and $V_{0}$ are the concentration and strength of disorder, respectively \cite{zeng2020fundamental}. In the case of our buckled honeycomb lattices, the $yxx$ component of nonlinear thermal Hall coefficient $\kappa^{BCD}$ can be written as

\begin{equation}
\kappa^{BCD} = \frac{7 \pi^3 k_{B}^{2} \lambda \Delta v_x v_y (\mu^2 - 2 \Delta^2)}{5 n \hbar^2 V_{0}^2 \mu^4 (\mu^2 + 3 \Delta^2)} k_B T^2.
\label{eq:nthkff}
\end{equation}

\begin{figure}
\centering
\includegraphics[scale=0.34]{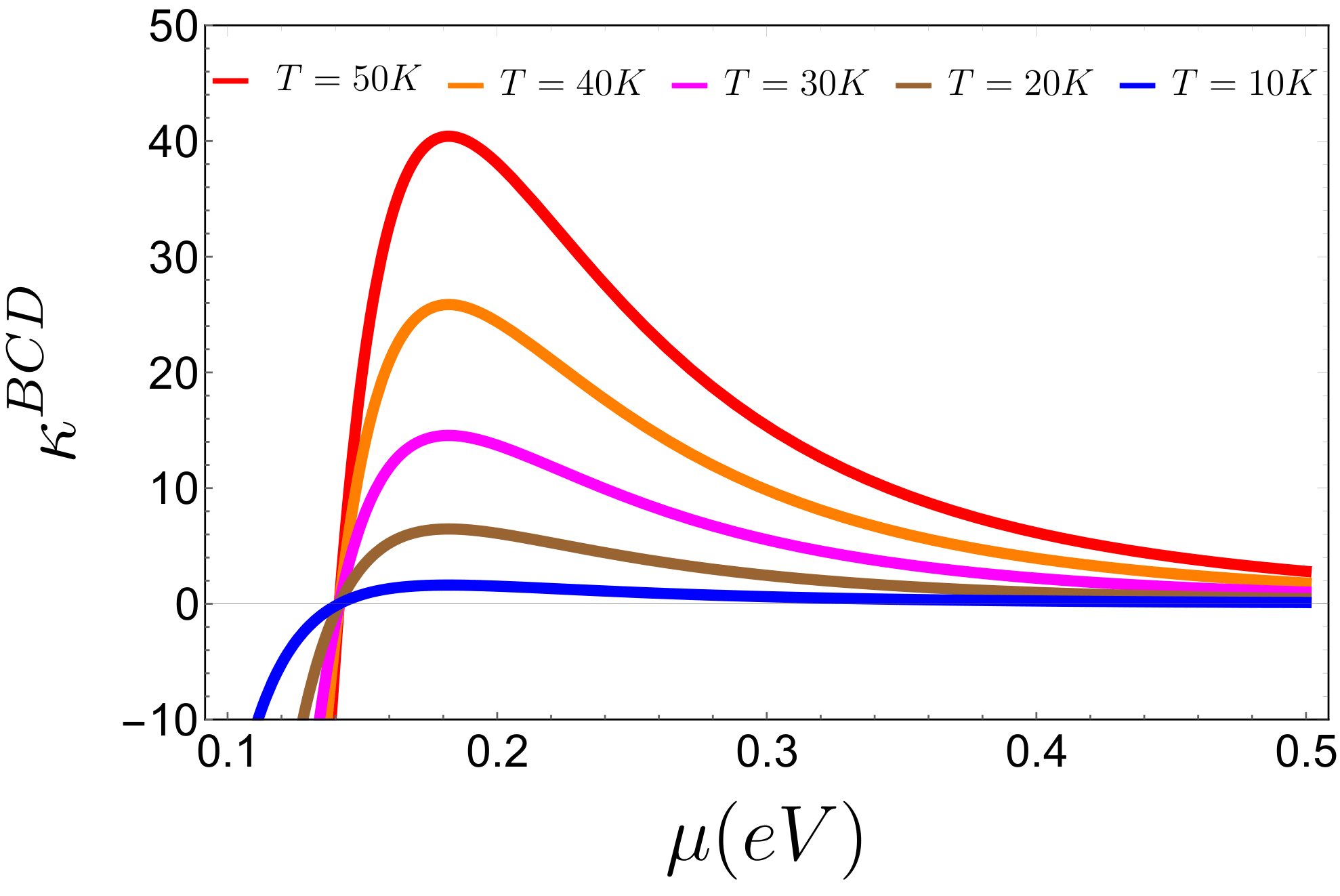}
\caption{\textbf{Nonlinear thermal Hall conductivity.} The variation of BCD induced nonlinear thermal Hall conductivity, $\kappa^{BCD}$, with chemical potential, $\mu$, for different temperature values. The $\kappa^{BCD}$ values are represented in the units of $10 ^{-2} k_{B}^{2}/\hbar$ \AA{}.}
\label{fig:thermal}
\end{figure}

We have studied the variation of the nonlinear Hall coefficient for the strained buckled honeycomb lattices (here, shown for stanene) with chemical potential, for different temperatures, presented in Fig.~\ref{fig:thermal}. We have considered representative values of the different parameters for our buckled honeycomb systems: $v_x = v_y = v_F = at = 6.11$ eV \AA{} (velocity for stanene given in Table \ref{tab:tabls}), $\Delta= 0.1$ eV (band gap of stanene given in Table \ref{tab:tabls}), $\lambda = 0.1 v$, $n V_0 = 100$ eV$^2$ \AA{}$^2$. Therefore, starting from the situation where the Fermi level touches the bottom of conduction band ($\mu \approx 0.1$ eV), the chemical potential can be tuned using a gate voltage. We find that for all temperatures, the nonlinear thermal Hall coefficient changes its sign and exhibits a peak at $\mu > 0.1$ eV for stanene. Further, the coefficient invariably goes to zero for large value of $\mu$. The peak value of nonlinear thermal Hall coefficient increases with increasing temperature. In a nutshell, BCD can give rise to the nonlinear thermal Hall conductivity, which varies quadratically in the temperature difference.

We note that a stronger BCD response can be attained by approaching the critical point with more precision. Finally, we compare the BCD response in the two distinct topological phases. All the strained buckled honeycomb lattices exhibit universal behaviour around the topological critical point. 
Similar to the observation during reversal of applied electric field, the sign of BCD changes when going from topologically non-trivial to trivial state. This is presented in Fig.~\ref{fig:4}(b). This can be understood to be the result of the exchange of Berry curvature between VB and CB around the topological phase transition. Overall, our results put forward a new platform to explore large and tunable BCD in buckled honeycomb lattices. This electrically switchable BCD will facilitate the exploration of various exotic quantum mechanical phenomena, such as NHE \cite{sodemann2015quantum}, chiral polaritonic effects \cite{basov2016polaritons}, nonlinear Nernst effect \cite{yu2019topological,zeng2019nonlinear}, nonlinear thermal Hall effect \cite{zeng2020fundamental}, and orbital-Edelstein \cite{yoda2018orbital} effects.

\section{Conclusions}

In summary, we introduced a class of elemental systems that exhibit electrically switchable giant BCD at the Fermi level. In particular, the elemental buckled honeycomb lattices -- silicene, germanene, and stanene -- exhibit an electric field-driven topological phase transition. The transverse electric field breaks the inversion symmetry of the systems and opens the possibility of obtaining a large BCD. However, the non-zero value of BCD is still restricted by the point group symmetry of the crystals. Therefore, we proposed that the symmetry of the crystals is reduced down to a single mirror plane using appropriate strain. The strain essentially perturbs the distribution of Berry curvature and induces asymmetry in a valley. Consequently, a sizable BCD is obtained for all the systems mentioned above. Moreover, a vanishing band gap near the critical band gap closing point triggers a giant BCD at the Fermi level. The value of BCD switches when the electric field strength crosses a critical value. This flipping can be explained in terms of the change in sign of Berry curvature across the critical point. Our study paves the way for exploring field tunable electrical and thermal nonlinear effects in a class of elemental systems.

\section{Acknowledgments}
A.B. sincerely acknowledges Indian Institute of Science for providing financial support. A.B. also wants to thank S. Roy, A. Bose, and A. Banerjee for illuminating discussions. N.B.J. acknowledges the support from the Prime Minister’s Research Fellowship. A.N. acknowledges support from the start-up grant (SG/MHRD-19-0001) of the Indian Institute of Science and DST-SERB (project number SRG/2020/000153).

\bibliographystyle{apsrev4-1} 
\bibliography{ref} 

\end{document}